\documentclass[12pt]{iopart}

\usepackage{graphicx}
\usepackage{epstopdf}
\usepackage{multibox}
\usepackage{dcolumn}

\usepackage{amssymb}
\usepackage{amstext}

\usepackage{hyperref}

\eqnobysec

\newcommand{\eqref}[1]{(\ref{#1})}

\def\d{\partial}

\newcommand{\dd}[2]{\frac{\d #1}{\d #2}}

\renewcommand{\exp}[1]{{\rm exp}(#1)}

\def\half{{\frac{1}{2}}}

\def\be{\begin{eqnarray}}
\def\ee{\end{eqnarray}}
\def\beann{\begin{eqnarray*}}
\def\eeann{\end{eqnarray*}}
\def\beq{\begin{equation}}
\def\eeq{\end{equation}}
\def\ba{\begin{array}}
\def\ea{\end{array}}
\def\ben{\begin{enumerate}}
\def\een{\end{enumerate}}
\def\bea{\begin{eqnarray}}
\def\eea{\end{eqnarray}}
\def\beann{\begin{eqnarray*}}
\def\eeann{\end{eqnarray*}}
\def\beq{\begin{equation}}
\def\eeq{\end{equation}}
\def\ba{\begin{array}}
\def\ea{\end{array}}
\def\ben{\begin{enumerate}}
\def\een{\end{enumerate}}

\def\5{\bar }
\def\6{\partial }
\def\7{\hat }
\def\4{\tilde }
\def\3#1{{{#1}^\prime}}

\def\cJ{{\cal J}}
\def\cK{{\cal K}}
\def\cL{{\cal L}}

\def\cP{{\cal P}}
\def\cQ{{\cal Q}}

\def\s0#1#2{\mbox{\small{$\frac{#1}{#2}$}}}

\def\ndelta{\delta\hspace{-0.50em}\slash\hspace{-0.05em} }

\begin{document}

\addtolength{\headsep}{4pt}

\begin{flushright}
{ULB-TH-06/08, \ gr-qc/0610130}
\end{flushright}

\title[Central charge for $\mathfrak{bms_3}$]{Classical
  central extension for asymptotic symmetries at null infinity in
  three spacetime dimensions}

\author{Glenn Barnich$^\ddagger$ and Geoffrey Comp\`ere$^\S$ }

\address{%
Physique Th\'eorique et Math\'ematique, Universit\'e Libre de
    Bruxelles and International Solvay Institutes, Campus
    Plaine C.P. 231, B-1050 Bruxelles, Belgium
} \ead{gbarnich@ulb.ac.be, gcompere@ulb.ac.be}

\begin{abstract} The symmetry algebra of asymptotically flat
  spacetimes at null infinity in three dimensions is the semi-direct
  sum of the infinitesimal diffeomorphisms on the circle with an
  abelian ideal of supertranslations. The associated charge algebra is
  shown to admit a non trivial classical central extension of Virasoro
  type closely related to that of the anti-de Sitter case.
\end{abstract}


\vspace{5cm} \footnotesize{$^\ddagger$Senior Research Associate of the
  National Fund for Scientific Research (Belgium)}

\footnotesize{$^\S$Research Fellow of the National Fund for Scientific
  Research (Belgium).}

\maketitle

\section{Introduction}

Classical central extensions in the Poisson bracket algebra
representation of asymptotic symmetries in general relativity have
been originally discovered in the example of three dimensional,
asymptotically anti-de Sitter spacetimes at spatial infinity
\cite{Brown:1986nw}. In this case, the Poisson algebra of surface
charges was shown to consist of two copies of the Virasoro
algebra. This fact has been relevant in the context of the ${\rm
  AdS}_3/{\rm CFT}_2$ correspondence \cite{Aharony:1999ti} and used to
give a microscopical derivation of the Bekenstein-Hawking entropy
for black holes with near horizon geometry that is locally ${\rm
AdS}_3$ \cite{Strominger:1997eq}. The analysis of the asymptotic
charge algebra has subsequently been performed in the context of
asymptotically de Sitter spacetimes at timelike infinity
\cite{Strominger:2001pn} with results very similar to those
obtained in the anti-de Sitter case.

For asymptotically flat-spacetimes, the appropriate boundary from
a conformal point of view is null infinity \cite{Witten:2001kn}.
The asymptotic symmetry algebra has been derived a long time ago
in four dimensions \cite{Bondi:1962px,Sachs:1962wk,Sachs:1962aa}
and more recently by conformal methods \cite{Penrose:1962ij} also
in three dimensions \cite{Ashtekar:1996cd}.

The purpose of this letter is to complete the picture for classical
central charges in three dimensions. We begin by computing the
symmetry algebra $\mathfrak{bms_n}$ of asymptotically flat spacetimes
at null infinity in $n$ dimensions, i.e., the $n$-dimensional analog
of the four dimensional Bondi-Metzner-Sachs algebra, by solving the
Killing equations to leading order.

In four dimensions, we make the obvious observation that the
asymptotic symmetry algebra can be larger than the one originally
discussed in \cite{Sachs:1962aa} if the conformal transformations of
the $2$-sphere are not required to be globally well-defined.  In three
dimensions, we recover the known results~\cite{Ashtekar:1996cd}:
$\mathfrak{bms_3}$ is the semi-direct sum of the infinitesimal
diffeomorphisms on the circle with the abelian ideal of
supertranslations.

In three dimensions, we then derive the space of allowed metrics
by requiring (i) that $\mathfrak{bms_3}$ be the symmetry algebra
for all allowed metrics, (ii) that the asymptotic symmetries leave
the space of allowed metrics invariant, (iii) that the associated
charges be linear and finite. As a new result, the associated
Poisson algebra of charges is shown to be centrally extended. A
non trivial central charge of Virasoro type with value
$c=\frac{3}{G}$ appears between the Poisson brackets of the
charges of the two summands. To conclude our analysis we point out
that the centrally extended asymptotic charge algebras in flat and
anti-de Sitter spacetimes are related in the same way than their
exact counterparts~\cite{Witten:1988hc}. Related recent work on
holography in asymptotically flat spacetimes can be found for
example in~\cite{referencesBMS}.

\section{Surface charges}

For pure Einstein gravity with a possibly non-vanishing cosmological
constant $\Lambda$, we associate to a vector $\xi$ the following $n-2$
forms in spacetime:
\begin{eqnarray}
  k_{\xi}[h,g]=- \delta_h k^K_\xi[g] -
  \xi \cdot \Theta[h,g] - k^S_{\cL_\xi g}[h,g], \label{eq:n-2forms}
\end{eqnarray}
where $\delta_h g_{\mu\nu}=h_{\mu\nu}$,
$(d^{n-p}x)_{\mu_1\dots\mu_p}\equiv \frac 1{p!(n-p)!}\,
\epsilon_{\mu_1\dots\mu_n} dx^{\mu_{p+1}}\dots dx^{\mu_n}$.
The first two terms of \eqref{eq:n-2forms} are expressed in the
form derived using covariant phase space methods
\cite{Iyer:1994ys,Wald:1999wa},
\begin{eqnarray}
k^K_\xi[g] = \frac{\sqrt{-g}}{16\pi G} (D^\mu \xi^\nu -  D^\nu
\xi^\mu )(d^{n-2}x)_{\mu\nu},\\
\Theta[h, g] = \frac{\sqrt{-g}}{16\pi G} (g^{\mu\alpha} D^\beta
h_{\alpha\beta} - g^{\alpha\beta} D^\mu h_{\alpha\beta})
(d^{n-1}x)_\mu.
\end{eqnarray}
The supplementary term
\begin{equation}
k^S_{\cL_\xi g}[h,g] = \frac{\sqrt{-g}}{16\pi G} (\half
g^{\mu\alpha} h_{\alpha\beta} (D^\beta \xi^\nu + D^\nu \xi^\beta)
- (\mu \leftrightarrow \nu) ) (d^{n-2}x)_{\mu\nu},\label{suppl}
\end{equation}
vanishes for exact Killing vectors of $g$, but not necessarily for
asymptotic ones. Expression \eqref{eq:n-2forms} has been derived using
cohomological methods in \cite{Barnich:1995db} and coincides with the
one originally derived in the context of four dimensional
asymptotically anti-de Sitter spacetimes \cite{Abbott:1981ff}.

By construction, the $n-2$ form $k_\xi$ is closed, $d k_\xi=0$, when
$g$ satisfies the equations of motion, $h$ the equations linearized
around $g$ and when $\xi$ is a Killing vector of $g$. Under these
conditions, its integral
\begin{eqnarray}
\ndelta Q_\xi[h,g]=\oint_{S}
k_\xi [h,g]\label{eq:21}
\end{eqnarray}
over a closed $n-2$ dimensional surface only depends on the homology
class of this surface. In the context of linearized gravity around a
solution $g$, one can furthermore prove
\begin{itemize}
\item for $n\geq 3$, that all $n-2$ forms that are closed for all
  solutions $h$ of the linearized theory are given, on solutions $h$,
  by $k_\xi$ for some Killing vector $\xi$, up to $d$ exact $n-2$
  forms which do not contribute to the integrals
  \cite{Barnich:1995db,Anderson:1996sc,Barnich:2004ts},

\item if $\xi_1,\xi_2$ are Killing vectors of the solution $g$ and $h$
  is a solution to the Einstein equations linearized around $g$, then
  so is $\cL_{\xi_1}h$ and \cite{Barnich:2001jy}
  \begin{eqnarray}
    \label{eq:22}
    \{\ndelta
    Q_{\xi_1},\ndelta Q_{\xi_2}\}[h,g]:= - \ndelta
    Q_{\xi_2}[\cL_{\xi_1}h,g]=
\ndelta Q_{[\xi_1,\xi_2]}[h,g].
  \end{eqnarray}

\item on solutions and for Killing vectors $\xi$ of $g$ with $S$
  chosen in the $t$ constant hyperplane, $\ndelta Q_\xi[h,g]$ coincides with
  the values computed in the context of the Hamiltonian formalism
  \cite{Regge:1974zd}.

\end{itemize}

The question is then how to use these exact results to get a theory of
asymptotic symmetries and charges in full general relativity for some
chosen boundary surface $S^\infty$ and background solution $\bar g$.
This involves two related aspects: finding the boundary conditions on
the fields and the asymptotic symmetry algebra. One possibility is to
impose boundary conditions such that, at the boundary, the theory of
charges for suitably defined asymptotic Killing vectors is controlled
by the linearized theory. We loosely refer to this situation as
``asymptotic linearity'' (cf.~\cite{Barnich:2001jy} for details). In
that case, one simply has
\begin{equation}
\oint_{S^\infty} k_\xi[h,g] = \oint_{S^\infty} k_\xi[h,\bar
g].\label{AL}
\end{equation}
and the charges of the full theory are defined by
\begin{eqnarray}
Q_\xi[g^f,\bar
g]=\oint_{S^\infty} k_\xi[g^f-\bar g,\bar g]+N_\xi[\bar g]\label{eq:2}.
\end{eqnarray}
While this approach trivially guarantees integrability of the charges,
it requires strong fall-off conditions that can be too restrictive and
eliminate solutions of interest. Indeed, these fall-off conditions
impose that the individual terms in the charges are finite, while in
general it may be necessary to allow for cancellation of infinite
contributions from different terms and for non linear corrections to
the charges (see e.g.~\cite{Henneaux:2002wm,Henneaux:2004zi}). Even
though these generalizations enlarge the space of allowed metrics, we
do not expect them to affect the conclusions concerning the central
charge below. This is the reason why we restrict ourselves at this
stage to the linear approach.

In the asymptotic linear case, the Poisson bracket of the charges is
given by $\{\cQ_{\xi_1},\cQ_{\xi_2}\}[g^f,\bar g]=-\delta^f_{\xi_1}
\cQ_{\xi_2}[g^f,\bar g]$, with $\delta_\xi g^f_{\mu\nu}=\cL_\xi
g^f_{\mu\nu}$ and
\begin{equation}
\{\cQ_{\xi_1},\cQ_{\xi_2}\}[g^f,\bar
g]=\cQ_{[\xi_1,\xi_2]}[g^f,\bar g]+\cK_{\xi_1,\xi_2}[\bar
g]-N_{[\xi_1,\xi_2]}[\bar g]. \end{equation}
It then follows that the expression
\begin{eqnarray}
\cK_{\xi_1,\xi_2}[\bar g] &=& \int_{S^\infty} k_{\xi_1}[\cL_{\xi_2} \bar g,\bar
g]\label{def_central}
\end{eqnarray}
defines a 2-cocycle on the Lie algebra of asymptotic Killing vector
fields $\xi$,
\begin{eqnarray}
\cK_{\xi_1,\xi_2}[\bar g]=-\cK_{\xi_2,\xi_1}[\bar g],\label{skewsym}\qquad
\cK_{[\xi_1,\xi_2],\xi_3}[\bar g] + \cK_{[\xi_2,\xi_3],\xi_1}[\bar g]
+ \cK_{[\xi_3,\xi_1],\xi_2}[\bar g]  = 0.\label{cocycle}
\end{eqnarray}
For Einstein gravity, the explicit formula for the central charge
follows from~\eqref{eq:n-2forms} and is given by\footnote{This
  expression differs from the one derived in \cite{Barnich:2001jy} by an
  overall sign because we have changed the sign convention for the
  charges and also by the fact that we use here the
  Misner-Thorne-Wheeler convention for the Riemann tensor. See also
  \cite{Koga:2001vq} for
  similar expressions.}
\begin{eqnarray}
\cK_{\xi,\xi^\prime}[\bar g]& =& \frac{1}{16\pi G }\oint_{S^\infty}
(d^{n-2}x)_{\mu\nu}\sqrt{- \bar g} \Big( - 2  \bar D_\sigma
\xi^\sigma \bar D^\nu \xi^{\prime\mu} + 2  \bar D_\sigma
\xi^{\prime\sigma} \bar D^\nu \xi^\mu \nonumber
 + 4  \bar D_\sigma \xi^\nu \bar D^\sigma \xi^{\prime \mu} \\
&& +\frac{8\Lambda}{2-n}\xi^\nu \xi^{\prime\mu} - 2 \bar R^{
\mu\nu\rho\sigma}\xi_\rho \xi_\sigma^\prime + (\bar D^\sigma
\xi^{\prime \nu}+ \bar D^\nu \xi^{\prime \sigma})(\bar D^\mu
\xi_\sigma + \bar D_\sigma \xi^\mu) \Big).\label{G_cbis}
\end{eqnarray}
Note that the central charge vanishes if either $\xi$ or $\xi^\prime$ is
an exact Killing vector of $\bar g$. Because \eqref{skewsym} can be
proved for all vector fields $\xi_1,\xi_2$ and solutions $g$ of Einstein's
equations, it follows from the definition of the Poisson
bracket that if either $\xi$ or $\xi^\prime$ is in addition an exact Killing
vector of $g^f$ then $\cQ_{[\xi,\xi^\prime]}[g^f,\bar g]=0$ by
choosing the normalization to vanish.

\section{The $\mathfrak{bms_n}$ algebra}

Introducing the retarded time $u=t-r$, the luminosity distance $r$ and
angles $\theta^A$ on the $n-2$ sphere by $x^1= r\cos \theta^1$, $x^A =
r\sin \theta^1\dots\sin \theta^{A-1}\cos \theta^A$, for
$A=2,\dots,n-2$, and
$x^{n-1}=r\sin \theta^1\dots \sin \theta^{n-2}$, the Minkowski metric
is given by
\begin{equation}
d\bar s^2=-du^2-2dudr
+r^2\sum_{A=1}^{n-2}s_A(d\theta^A)^2,\label{metric_flat}
\end{equation}
where $s_1 = 1$, $s_A = \sin^2 \theta^1\dots \sin^2\theta^{A-1}$ for
$2\leq A\leq n-2$. The (future) null boundary is defined by $r
=constant\rightarrow \infty$ with $u,\theta^A$
held fixed.

We require asymptotic Killing vectors to satisfy the Killing equation
to leading order. They have the form $\xi^\mu= \chi^\mu
\tilde\xi^{(\mu)}(u,\theta)+o(\chi^\mu)$ for some fall-offs
$\chi^\mu(r)$ to be determined. Here, round brackets around a single
index mean that the summation convention is suspended. For such
vectors, $\cL_\xi \bar g_{\mu\nu}= O(\rho_{\mu\nu})$. Solving the
Killing equation to leading order means finding the highest orders
$\chi^{\mu}(r)$ in $r$ such that equation
\begin{equation}
\cL_\xi \bar g_{\mu\nu} = o(\rho_{\mu\nu}),\label{def_akV2}
\end{equation}
admits non-vanishing $\tilde \xi^{\mu}(u,\theta)$ as solutions. After a
straightforward computation (summarized in \ref{app:aKv_known}), one
finds
\begin{equation}\eqalign{
\xi^u&= T(\theta^A)+u\d_{1}Y^1(\theta^A)+o(r^0),\qquad
\xi^r= -r\d_{1} Y^1(\theta^A)+o(r),\\
\xi^A&= Y^A(\theta^B)+o(r^0),\qquad A=1\dots n-2.}\label{asykvf}
\end{equation}
where $T(\theta^A)$ is an arbitrary function on the $n-2$ sphere,
and $Y^A(\theta^A)$ are the components of the conformal Killing
vectors on the $n-2$ sphere. These asymptotic Killing vectors form a
sub-algebra of the Lie algebra of vector fields and the bracket
induced by the Lie bracket $\hat \xi=[\xi,\xi^\prime]$ is determined by
\begin{eqnarray}
\hat T= Y^A\d_{A} T^\prime + T \d_{1} Y^{\prime 1} - Y^{\prime
A}\d_{A} T -T^\prime \d_{1} Y^{ 1}\,, \\
\hat Y^A = Y^B\d_{B} Y^{\prime A} -Y^{\prime B}\d_{B} Y^{A}.\label{bracket}
\end{eqnarray}
It follows that asymptotic Killing vectors with $T=0=Y^A$ form an
ideal in the algebra of asymptotic Killing vectors. The quotient
algebra is defined to be $\mathfrak{bms_n}$.  It is the semi-direct
sum of the conformal Killing vectors $Y^A$ of Euclidean $n-2$
dimensional space with an abelian ideal of so-called
infinitesimal supertranslations. Note that the exact Killing vectors
of $\bar g$, $\xi_\mu =a_\mu+b_{[\mu\nu]}x^\nu$ give rise to
\begin{eqnarray}
Y^A_E=\frac{1}{
  s_{(A)}}(b_{[i0]}+b_{[ij]}\frac{x^j}{r})\frac{1}{r}\dd{x^i}{y^{A}},\
T_E=-[a_0+a_i\frac{x^i}{r}],\label{eq:1}
\end{eqnarray}
and belong to $\mathfrak{bms_n}$, so that $\mathfrak{iso(n-1,1)}$ is a
subalgebra of $\mathfrak{bms_n}$.

In order to make contact with conformal methods, we just note that if
$\tilde g_{\mu\nu} = r^{-2}\bar g_{\mu\nu}$ is the metric induced at
the boundary $r$ constant,
\begin{equation}
d \tilde s^2= -\frac{1}{r^2} \, du^2+
\sum_{A=1}^{n-2}s_A(d\theta^A)^2.\label{B}
\end{equation}
one can easily verify that $\mathfrak{bms_n}$ is isomorphic to the Lie
algebra of conformal Killing vectors of the boundary metric~\eqref{B},
in the limit $r\rightarrow \infty$.

For $n> 4$, the asymptotic algebra contains the infinitesimal
supertranslations parameterized by $T(\theta^A)$ and the $n(n-1)/2$
dimensional conformal algebra of Euclidean space
$\mathfrak{so(n-1,1)}$ in $n-2$ dimensions, isomorphic to the Lorentz
algebra in $n$ dimensions.

In four dimensions, the conformal algebra of the $2$-sphere is
infinite-dimensional and contains the Lorentz algebra
$\mathfrak{so(3,1)}$ as a subalgebra. It would of course be
interesting to analyze whether central extensions arise in the charge
algebra representation of $\mathfrak{bms_4}$, but we will not do so
here. Note that in the original discussion \cite{Sachs:1962aa}, the
transformations were required to be well-defined on the $2$-sphere and
$\mathfrak{bms_4}$ was restricted to the semi-direct sum of
$\mathfrak{so(3,1)}$ with the infinitesimal supertranslations. In this
case, there are no non trivial central extensions, see
e.g.~\cite{mccarthy:1978}.

In three dimensions, the conformal Killing equation on the circle
imposes no restrictions on the function $Y(\theta)$. Therefore,
$\mathfrak{bms_3}$ is characterized by $2$ arbitrary functions
$T(\theta),Y(\theta)$ on the circle. These functions can be Fourier
analyzed by defining $P_n \equiv \xi(T=\,\exp{i n \theta},Y=0)$ and
$J_n = \xi(T=0,Y=\exp{in\theta})$. In terms of these generators, the
commutation relations of $\mathfrak{bms_3}$ become
\begin{equation}
i[ J_m,J_n] = (m-n) J_{m+n},\qquad  i[ P_m,P_n] =
0,\label{alg_bms}\qquad  i[J_m,P_n] = (m-n) P_{m+n}.
\end{equation}
In other words, the 6 dimensional Poincar\'e algebra
$\mathfrak{iso(2,1)}$ of 3 dimensional Minkowski spacetime is enhanced
to the semi-direct sum of the infinitesimal diffeomorphisms on the
circle with the infinitesimal supertranslations.

\section{Charge algebra representation of  $\mathfrak{bms_3}$}

In order to determine the Poisson algebra representation of
$\mathfrak{bms_3}$ we need to specify the boundary conditions on the
metric, $g_{\mu\nu} = \bar g_{\mu\nu}+h_{\mu\nu}$, where $h_{\mu\nu} =
\chi_{\mu\nu}(r)$ are fall-offs to be determined, and also subleading
terms in the asymptotic Killing vectors.

If we want the asymptotic symmetry algebra to be the same for all
allowed metrics, we need to require that solving the Killing equation
to leading order for $g$ in place of $\bar g$ will lead to the
$\xi^\mu$ given in~(\ref{asykvf}). We will also need $\cL_\xi
g_{\mu\nu} = O(\chi_{\mu\nu})$ so that the asymptotic symmetry
$\cL_\xi g_{\mu\nu}$ leaves the space of allowed metrics
invariant. These conditions are satisfied for metric deviations of the
form
\begin{eqnarray}
h_{uu} = O(1), \qquad h_{ur} = O(r^{-1}), \qquad h_{u\theta} =
O(1),\nonumber\\
h_{rr} = O(r^{-2}),\qquad h_{r\theta} = O(1), \qquad
h_{\theta\theta} = O(r),\label{BC_bms}
\end{eqnarray}
and asymptotic Killing vectors defined by
\begin{eqnarray}
\xi^u&=& T(\theta)+u\d_{\theta}Y(\theta)+O(r^{-1}),\nonumber\qquad
\xi^r=-r\d_{\theta} Y(\theta)+O(r^0),\\
\xi^\theta&=&Y(\theta)-
\frac{u}{r}\d_\theta\d_{\theta}Y^{\theta}(\theta)+
\frac{1}{r}f_{sub}^{\theta}(\theta)+O(r^{-2}),\label{bms_complete}
\end{eqnarray}
where $f^\theta_{sub}(\theta)$ is an arbitrary function. In addition, the
associated charges are finite and linear in the sense of condition  (\ref{AL})
for metric deviations of the form
\begin{eqnarray}
h_{uu} = O(1), \qquad h_{ur} = O(r^{-1}), \qquad h_{u\theta}
=O(1),\nonumber\\
h_{rr} = O(r^{-2}),\qquad h_{r\theta} = h_1(\theta) + O(r^{-1}),
\qquad h_{\theta\theta} =  h_{2}(\theta) r
+O(r^0).\label{BC_bms_bis}
\end{eqnarray}
These boundary conditions contain for example the metric
\begin{equation}
ds^2 = -(1-4m)^2du^2-2 du dr - 8J(1-4m)dud\theta -
\frac{8J}{1-4m}drd\theta+(r^2-16J^2)d\theta^2,\label{kerr3}
\end{equation}
which describes a spinning particle in Minkowski spacetime
\cite{Deser:1983tn}. The space of allowed metrics also contains the
dimensional reduction of the Einstein-Rosen waves from four to three
dimensions \cite{Ashtekar:1996cm}, for which the metric deviation at
infinity in a suitable coordinate system is given by $h_{uu} = O(1)$,
$h_{ur}=O(r^{-1})$, the others zero.

The charges (\ref{eq:2}) reduce to
\begin{eqnarray}
\hspace*{-2cm} Q_\xi[g^f,\bar
g]=\frac{1}{16\pi G}\int_{0}^{2\pi}d\theta\,\Big(
h_{uu}T+\big(2h_{u\theta}-u\d_\theta h_{uu} +2\d_\theta\d_\theta h_1
+\d_\theta h_2 +r\d_u h_{r\theta}-r\d_\theta h_{ur}\big)Y \Big), \label{eq:3}
\end{eqnarray}
where the charge of the background has been set to zero. These charges
obviously vanish for trivial asymptotic Killing vectors. One can also
verify that when the metric deviations satisfy the linearized Einstein
equations these charges are conserved, i.e., independent of
$u$. Finally, we note that $\int_{0}^{2\pi}\,d\theta\, k^S_{\cL_\xi
g}[h,g]=0$ for $\xi,h$ satisfying (\ref{bms_complete}) respectively
(\ref{BC_bms_bis}) so that the charges agree with those defined in
\cite{Iyer:1994ys}.

The expression \eqref{eq:3} allows us to compute
the central extension of the Poisson algebra representation of
$\mathfrak{bms_3}$ by replacing $h_{\mu\nu}$ by $\cL_{\xi^\prime} \bar
g_{\mu\nu}$ with $\xi^\prime$ given in (\ref{asykvf}). The result is
\begin{eqnarray}
\cK_{\xi,\xi^\prime} &=& \frac{1}{8\pi G}\int_0^{2\pi} d\theta
 \Big[\d_\theta Y^\theta (\d_\theta\d_\theta T^\prime + T^\prime) -
\d_\theta Y^{\prime\theta} (\d_\theta\d_\theta T + T)
 \Big].
\end{eqnarray}
Using algebra (\ref{bracket}), cocyle condition (\ref{cocycle}) can be
explicitly checked. In terms of the generators $\cQ_{P_n} = \cP_n,
\cQ_{J_n} = \cJ_n$, we get the centrally extended algebra
\begin{eqnarray}
i\{ \cJ_m,\cJ_n \} = (m-n) \cJ_{m+n},\qquad
i\{ \cP_m,\cP_n \} = 0,\nonumber \\
i\{ \cJ_m,\cP_n \} = (m-n) \cP_{m+n}+\frac{1}{4G}
m(m^2-1)\delta_{n+m,0}.\label{bms_charge}
\end{eqnarray}
It can easily be shown to be non-trivial in the sense that it cannot
be absorbed into a redefinition of the generators. Only the
commutators of generators involving either $\cJ_0,\cJ_1, \cJ_{-1}$ or
$\cP_0,\cP_1,\cP_{-1}$ corresponding to the exact Killing vectors of
the Poincar\'e algebra $\mathfrak{iso(2,1)}$ are free of central
extensions.

The algebra~\eqref{bms_charge} has many features in common with the
anti-de Sitter case: it has the same number of generators, and a
Virasoro type central charge. In fact, these algebras are related in
the same way than their exact counterparts \cite{Witten:1988hc}: if
one introduces the negative cosmological constant $\Lambda =
-\frac{1}{l^2}$ and considers
\begin{equation}
\hspace{-1cm} i[ {J_m},{J_n}] = (m-n) J_{m+n},\quad i[ {P_m},{P_n}
] = \frac{1}{l^2} (m-n)J_{m+n},\label{alg_bms2}\quad i[{J_m},{P_n}
] = (m-n) P_{m+n},
\end{equation}
the $\mathfrak{bms}_3$ algebra~\eqref{alg_bms} corresponds to the case
$l\rightarrow \infty$. For finite $l$, the charges
$\cL_m^\pm$ corresponding to the generators $L_m^{\pm} = \half (\, l
P_{\pm m}
\pm J_{\pm m})$ form the standard two copies of the
Virasoro algebra,
\begin{eqnarray}
{i} \{ \cL^\pm_m,\cL^\pm_n \} &=&
(m-n) \cL^\pm_{m+n}
+\frac{c}{12}m(m^2-1)\delta_{n+m,0} ,\qquad
\{ \cL^\pm_m,\cL^\mp_n \} =\, 0,
\end{eqnarray}
where $c= \frac{3l}{2G}$ is the central charge for
  the anti-de Sitter case.

\ack

The authors are grateful to S.~Detournay, J.~Gegenberg, G.~Giribet,
C.~Mart\'{\i}nez, P.~Spindel and R.~Troncoso for useful
discussions. This work is supported in part by a ``P{\^o}le
d'Attraction Interuniversitaire'' (Belgium), by IISN-Belgium,
convention 4.4505.86, by the National Fund for Scientific Research
(FNRS Belgium) and by the European Commission programme
MRTN-CT-2004-005104, in which the authors are associated to
V.U.~Brussels.

\appendix

\section{Explicit computation of the $\mathfrak{bms}_n$ algebra}
\label{app:aKv_known}

Introducing the notation $\tilde\xi^u=U(u,\theta^A)$,
$\tilde\xi^r=R(u,\theta^A)$, $\tilde \xi^A=Y^A(u,\theta^B)$, the
$rr$-component of equation~\eqref{def_akV2} reduces to
\begin{equation}
-2 U \d_r\chi^u+\d_r o(\chi^u) =o(\rho_{rr}).
\end{equation}
The equation requires $\chi^u=r^0$. Since $\d_r o(r^0) =
o(r^{-1})$, we have $\rho_{rr} = r^{-1}$. The $ur$-component of
equation~\eqref{def_akV2} then gives
\begin{equation}
-\d_u U +o(r^{0}) -\d_r\chi^r R + \d_r o(\chi^r) =o(\rho_{ur}).
\end{equation}
This leads to $\chi^r = r$, $R + \d_u U = 0$ and $\rho_{ur}=r^0$.
The $uu$-component of equation~\eqref{def_akV2} reduces to
\begin{equation}
- 2 r \d_u R + o(r) =o(\rho_{uu}).
\end{equation}
It imposes $\d_u R = 0$ and gives $\rho_{uu} = r$. From the $rA$
component,
\begin{eqnarray}
  \label{eq:11}
  -\partial_{A} U+o(r^0)+ \partial_r \chi^A Y^A r^2 s_A
+r^2\partial_r  o(\chi^A)=o(\rho_{rA}),
\end{eqnarray}
we get $\chi^A=r^0$ and $\rho_{rA}=r$. The $uA$-component of
equation~\eqref{def_akV2} is
\begin{eqnarray}
r^2 \d_u Y^A + r^2 o(r^0) +r\d_A R + o(r^1) &=&o(\rho_{uA}),
\end{eqnarray}
implying $\d_u Y^A = 0$, and $\rho_{uA}=r^2$. Finally, the $AA$
and $AB$ with $A\neq B$ components of equation~\eqref{def_akV2}
are given by
\begin{eqnarray}
2 r^2 R s_{A} + 2r^2 \d_{(A)} Y^{(A)} s_{(A)} + r^2
Y^C \d_C s_{A} +o(r^2)&=& o(\rho_{AA}),\\
 r^2\d_B Y^{(A)} s_{(A)} + r^2 \d_A Y^{(B)}s_{(B)}+o(r^2) &=&
o(\rho_{AB}).
\end{eqnarray}
One finds the following conditions
\begin{eqnarray}
\fl\d_u Y^A = 0, \quad R + \d_A Y^{(A)} + \sum_{C<A}Y^C
\cot{\theta^C}=0,\quad \d_B Y^{(A)} s_{(A)} + \d_A Y^{(B)}s_{(B)}
= 0,
\end{eqnarray}
with $\rho_{AA}=r^2=\rho_{AB}$. The constraints imposed by
\eqref{def_akV2} on $U$, $R$ and $Y^A$ are summarized by
\begin{eqnarray}
&R = -\d_1 Y^1, \label{R},\qquad \d_u U=\d_{1}Y^1,\qquad\d_u \d_u U = 0,\qquad \d_u Y^A = 0,\label {U}\\
 & \d_{1} Y^1=\d_{{(A)}}Y^A+\sum_{B<A}\cot \theta^B Y^B,\quad \forall A,
\label{eq:conf1}\\
&\d_{A} Y^{B}s_{(B)}+\d_{B} Y^{A}s_{(A)}=0,\quad A\neq B,\quad
A,B=1,\dots,n-2. \label{eq:conf2}
\end{eqnarray}
The last two equations allow one to identify $Y^A(\theta^B)$ with
the conformal Killing vectors of the sphere in $n-2$ dimensions
with metric $g^{(n-2)}_{AB} = \delta_{AB}s_{(A)}$.

\section*{References}

\bibliography{/Users/gbarnich/Documents/Physics/Bibliography/master2}

\providecommand{\href}[2]{#2}\begingroup\raggedright\endgroup

\end{document}